\documentclass[modern,manuscript]{aastex61}

\usepackage{rotating}
\usepackage{mathtools}
\usepackage[graphicx]{realboxes}

\newcommand{\lalpha}{Ly$\alpha$}
\newcommand{\pimenc}{$\pi$ Men c}
\newcommand{\pimenb}{$\pi$ Men b}
\newcommand{\pimen}{$\pi$ Men}

\received{????, 2019}
\submitjournal{ApJL}

\shorttitle{{\pimenc} and the characterization of small worlds}
\shortauthors{Garc\'ia Mu\~noz et al.}

\begin{document}

\title{Is $\pi$ Men c's atmosphere hydrogen-dominated? \\
Insights from a non-detection of H {\sc i} {\lalpha} absorption.}

\correspondingauthor{Antonio Garc\'ia Mu\~noz}
\email{garciamunoz@astro.physik.tu-berlin.de, tonhingm@gmail.com}

\author{A. Garc\'ia Mu\~noz}
\affiliation{Zentrum f\"ur Astronomie und Astrophysik, Technische Universit\"at Berlin, 
Hardenbergstrasse 36, D-10623, Berlin, Germany}

\author{A. Youngblood}
\affiliation{NASA Goddard Space Flight Center, 
Greenbelt, MD 20771, USA}
\affiliation{
Laboratory for Atmospheric and Space Physics, 1234 Innovation Drive, Boulder, CO 80303, USA}

\author{L. Fossati}
\affiliation{Space Research Institute, Austrian Academy of Sciences,
Schmiedlstrasse 6, A-8042 Graz, Austria}

\author{D. Gandolfi}
\affiliation{
Dipartimento di Fisica, Universit\`a degli Studi di Torino, 
via Pietro Giuria 1, I-10125, Torino, Italy
}

\author{J. Cabrera}
\affiliation{
Deutsches Zentrum f\"ur Luft- und Raumfahrt, 
Institut f\"ur Planetenforschung, D-12489 Berlin, Germany
}

\author{H. Rauer}
\affiliation{
Deutsches Zentrum f\"ur Luft- und Raumfahrt, 
Institut f\"ur Planetenforschung, D-12489 Berlin, Germany
}
\affiliation{Zentrum f\"ur Astronomie und Astrophysik, Technische Universit\"at Berlin, 
Hardenbergstrasse 36, D-10623, Berlin, Germany}
\affiliation{
Institute of Geological Sciences, Freie Universit\"at Berlin, 
Malteserstrasse 74-100, D-12249, Berlin, Germany  
}

\begin{abstract}
\newpage

Constraining the composition of super-Earth-to-sub-Neptune-size
planets is a priority to understand the processes of planetary formation and evolution. 
{\pimenc} represents a unique target for the atmospheric and compositional   
characterization of such planets because 
it is strongly irradiated  
and its bulk density is consistent with abundant H$_2$O. 
We searched for hydrogen from photodissociating
H$_2$/H$_2$O  in {\pimenc}'s upper atmosphere through 
 H {\sc i} {\lalpha} transmission spectroscopy 
with the Hubble Space Telescope's STIS instrument, but did not detect it. 
We set 1$\sigma$ (3$\sigma$) upper limits  for the effective planet-to-star size ratio 
$R_{Ly\alpha}$/$R_{\star}$=0.13 (0.24) and 0.12 (0.20) at velocities 
[$-$215,$-$91] km/s and [$+$57,$+$180] km/s, respectively. 
We reconstructed the stellar spectrum, and estimate that 
{\pimenc} receives about 1350 erg cm$^{-2}$ s$^{-1}$ of 5--912-{\AA}-energy, 
enough to cause rapid atmospheric escape. 
An interesting scenario to explain the non-detection 
is that {\pimenc}'s atmosphere is dominated by H$_2$O 
or other heavy molecules rather than H$_2$/He.
According to our models, 
abundant oxygen results in less extended atmospheres, 
which transition from neutral to ionized hydrogen 
closer to the planet. 
We compare our non-detection to other detection attempts, 
and tentatively identify two behaviors: 
planets with densities $\lesssim$2 g cm$^{-3}$ (and likely hydrogen-dominated atmospheres)
result in H {\sc i} {\lalpha} absorption, 
whereas planets with densities $\gtrsim$3 g cm$^{-3}$ (and plausibly 
non-hydrogen-dominated atmospheres) do not result in measurable absorption. 
Investigating a sample of strongly-irradiated sub-Neptunes
may provide some statistical confirmation if it is shown that they do not
generally develop extended atmospheres.
\\

\end{abstract}


\section{Introduction} \label{sec:intro}

Planets in the super-Earth-to-sub-Neptune size range
(radii 1$<$$R_{\rm{p}}$/$R_{\earth}$$<$3.8)
are not represented in our 
Solar System but are numerous around other stars 
\citep{batalha2014,marcyetal2014}. 
Yet, key aspects such as their composition or the mechanisms controlling their formation 
and evolution remain unclear.
This range of sizes overlaps with the transition between rocky planets and planets with 
large amounts of gases and astrophysical ices \citep{valenciaetal2013,rogers2015}. 
Understanding this transition is critical towards forming the big picture of exoplanets,   
and as a preparatory step to investigate planets with conditions apt for life. 
Determining the atmospheric composition of a sample of small exoplanets will prove
useful to address these and other open questions in exoplanetary science.
As small exoplanets are notoriously difficult to characterize, 
space missions such as CHEOPS \citep{fortieretal2014}, TESS \citep{rickeretal2015} and 
PLATO \citep{raueretal2014} will play a key role at finding the most favorable targets
around nearby stars for follow-up investigations.
\\

{\pimen} is a quiet G0 V star of V-mag=5.65 at 18.28 pc \citep{gaiadr22018}
 orbited by two substellar objects. 
{\pimenb} is a massive ($M_{\rm{p}}$/$M_{\rm{J}}$$\sim$10) object on an eccentric 
($e$$\sim$0.6), long-period ($P_{\rm{orb}}$$\sim$2100 days) orbit 
discovered by radial velocity \citep{jonesetal2002}. 
{\pimenc}, 
a small ($M_{\rm{p}}$/$M_{\earth}$=4.52$\pm$0.81; 
$R_{\rm{p}}$/$R_{\earth}$=2.06$\pm$0.03), close-in ($P_{\rm{orb}}$=6.27 days) 
planet \citep{gandolfietal2018}, 
 was recently discovered using transit photometry with TESS
\citep{gandolfietal2018, huangetal2018}. 
Its bulk density ($\rho_{\rm{p}}$=2.82$\pm$0.53 g cm$^{-3}$) places it in a 
region of the mass-radius diagram consistent with a variety of 
compositions ranging from 100\% water to 
a rocky core surrounded by small amounts by mass of hydrogen-helium. 
Interestingly, the planet sits near the radius gap 
($R_{\rm{gap}}$/$R_{\earth}$$\sim$1.8) of small, close-in planets 
\citep{fultonetal2017,vaneylenetal2018} 
that separates the planets that are thought
to have retained an atmosphere ($R_{\rm{p}}$/$R_{\rm{gap}}$$>$1)
from those that lost it as a result of 
irradiation-driven escape ($R_{\rm{p}}$/$R_{\rm{gap}}$$<$1) 
\citep{lopezfortney2013,owenwu2013}. 
These arguments strongly suggest that 
{\pimenc} has an atmosphere that is currently escaping. 
\\

We report the first attempt to detect {\pimenc}'s upper atmosphere with Hubble 
Space Telescope (HST) observations of H {\sc i} {\lalpha} absorption. 
The search is motivated by the planet's bulk density, consistent with  
large amounts of hydrogen in the form of H$_2$ or H$_2$O. 
Under the effects of strong stellar irradiation, photodissociation of both molecules 
will produce H atoms that will escape the planet 
and form a potentially detectable extended atmosphere. 
The same strategy has revealed the occurrence of significant H {\sc i} {\lalpha} absorption
in the atmospheres of the hot Jupiters HD 209458 b and HD 189733 b 
\citep{vidalmadjaretal2003,vidalmadjaretal2004,
benjaffel2007,benjaffel2008,bourrieretal2013, 
 lecavelierdesetangsetal2010,lecavelierdesetangsetal2012},  
and the warm Neptunes GJ 436 b and GJ 3470 b 
\citep{kulowetal2014,ehrenreichetal2015,bourrieretal2018}. 
The attempts to detect H {\sc i} {\lalpha} absorption around smaller and/or less 
irradiated planets 
have so far resulted in non-detections
or in less clear conclusions: 
HD 97658 b \citep{bourrieretal2017a}, 
55 Cancri e \citep{ehrenreichetal2012}, 
TRAPPIST-1 b and c \citep{bourrieretal2017b,bourrieretal2017c}, 
Kepler-444A e and f \citep{bourrieretal2017d}, and 
GJ 1132 b \citep{waalkesetal2019}. 
The theoretical understanding of when planets develop extended atmospheres is 
imperfect, 
and in particular the role played by atmospheric composition remains poorly explored.
\\

\section{Reconstructed Spectrum of $\pi$\,Men} \label{sec:spectrum}

\begin{figure}
    \centering
    \includegraphics[width=\textwidth]{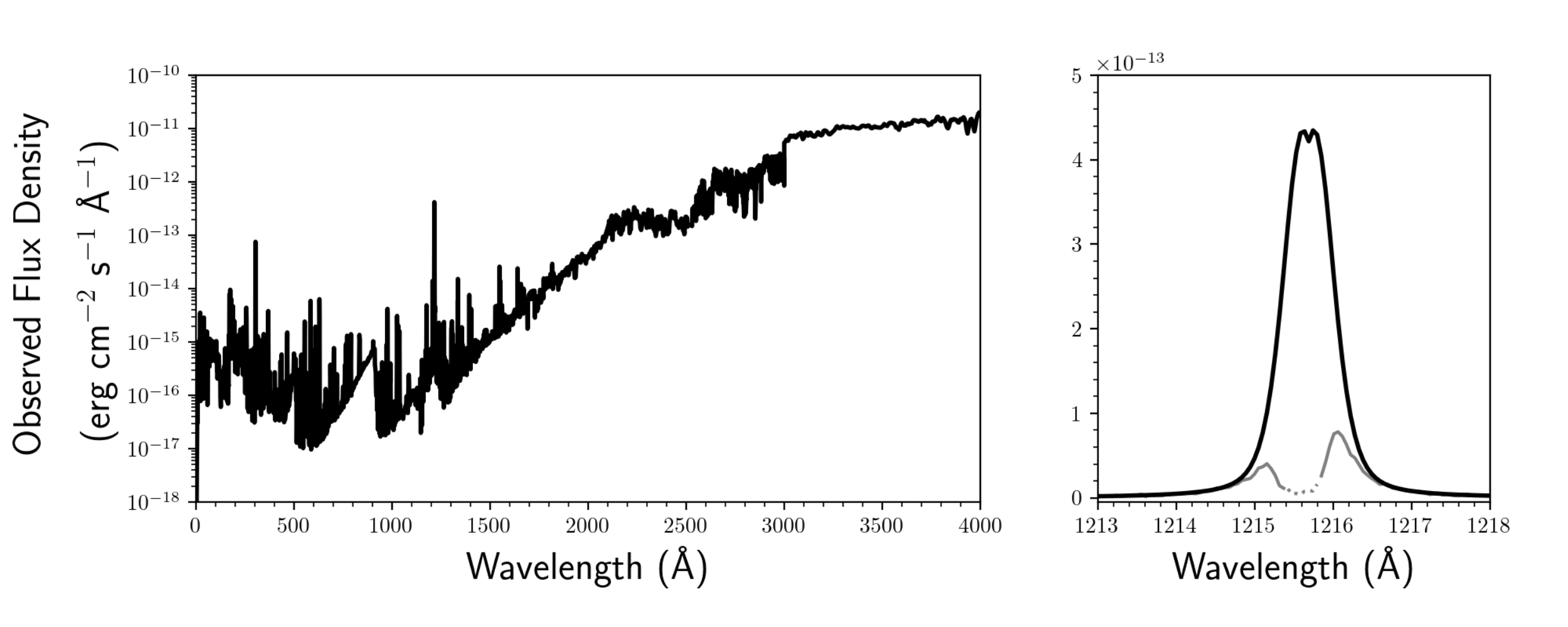}
    \caption{\textit{Left:} The reconstructed 5-4000 {\AA} spectrum of {\pimen} is shown here  
    as would be observed at 18.28 pc. 
    \textit{Right:} 
    The coadded STIS G140M spectrum is shown in grey and the reconstructed intrinsic 
    Ly$\alpha$~profile is shown in black, 
    convolved to the instrument spectral resolution.
    The dotted line shows the geocoronal airglow-contaminated region.}
    \label{fig:SED}
\end{figure}

Our hydrodynamic-photochemical model, which is the basis for the interpretation of the
reported observations (see below), requires a realistic stellar spectrum 
as input into the top of the atmosphere. 
To this end, we constructed a complete 5 \AA--2.5 {$\mu$}m spectrum for {\pimen} 
from new and archival data, scaled solar spectra, and a radiative equilibrium 
stellar model. 
Table~\ref{table:data_sources} gives the source for each part of the reference
spectrum including any scaling factors used. 
The left panel of Fig.~\ref{fig:SED} shows the representative spectrum of {\pimen} as observed
from Earth.
In the rest of this section we describe the reconstruction of the Ly$\alpha$~line, 
which is based on new data from this work obtained to search for a planetary transit signature.\\

We observed {\pimen} on 2019 Jul 24 during 5 consecutive orbits with the Hubble Space Telescope's (HST) 
STIS instrument as part of HST-GO-15699. The first two orbits occurred during pre-transit, 
the following two during transit, and the final orbit occurred post-transit, with respect 
to the transit observed by TESS. Our data were taken with the G140M grating centered at 
1222\,\AA\ and the 52$\times$0.2\arcsec\ slit, in time-tag mode. 
The complete spectral range covered by each spectrum is 1194--1249\,\AA. 
We downloaded from the MAST archive the data calibrated and extracted by {\it calstis}\footnote{{\tt http://www.stsci.edu/hst/stis/software/analyzing/calibration/pipe\_soft\_
hist/intro.html}}.\\

To reconstruct the bright H {\sc i} Ly$\alpha$ line (1216\,\AA) from the STIS spectrum, 
we used methods described in \citet{youngbloodetal2016} to simultaneously fit a model of the 
ISM H {\sc i} and D {\sc i} absorption and a model of the intrinsic stellar emission. 
Given the non-detection of the planetary transit (see below), 
we co-added the spectra obtained in each of the 5 HST orbits for improved S/N. 
In contrast to the procedure described in \citet{youngbloodetal2016}, we assumed a Voigt profile 
for the intrinsic stellar emission, and a small Gaussian in absorption to account for the 
self-reversal of the line expected in G dwarfs like $\pi$~Men. We find an intrinsic 
Ly$\alpha$ flux of F(Ly$\alpha$)\,=\,3.80$\pm$0.40\,$\times$\,10$^{-13}$\,erg\,cm$^{-2}$s$^{-1}$\,. 
The right panel of Fig.~\ref{fig:SED} shows the reconstruction compared to the STIS data. 
For the ISM, the best fitting parameters are $\log{N(HI)}$\,=\,18.51$\pm$0.02\,cm$^{-2}$, $b$\,=\,12.5$\pm$0.7\,km/s, 
and $v_{HI}$\,=\,-7.4$^{+0.7}_{-0.8}$\,km/s. Accounting for a systemic offset of -7.25 km/s 
in the STIS wavelength solution, we find that our fitted stellar radial velocity 
(+3.48$^{+0.76}_{-0.74}$ km/s) and ISM radial velocity (-7.4$^{+0.7}_{-0.8}$\,km/s) agree 
well with measurements from Gaia DR2 \citep{gaiadr22018} (+10.73\,km/s) and predictions from 
the ISM Kinematic Calculator from \citet{redfieldlinsky2008} (G and Vel clouds have velocities of 
$-$2.26$\pm$1.29\,km/s and +2.28$\pm$0.76\,km/s, respectively). 
\\

We note that the {\lalpha} flux levels detected in our STIS spectrum are less than expected, 
because of the unknown ISM H {\sc i} column density, Doppler broadening parameter, 
and radial velocity at the 
time of planning these observations. The relative velocity between the ISM absorption 
and stellar emission is -11 km/s, which inconveniently makes the blue wing of the observed 
{\lalpha}~profile, where the strongest escaping atmosphere signatures are expected, 
more strongly attenuated than the red wing. 
Also, our fitted H {\sc i} column density, while not atypical for a star at 18 pc, 
is on the upper end of the expected range. A priori knowledge of the ISM column density 
as a function of celestial coordinates would be helpful in planning exoplanet transit 
observations using ISM-affected lines as a backlight.
For reference, Fig. \ref{fig:Lya_Gdwarfs} of Appendix \ref{ref:Gdwarfs} compares the
{\lalpha} line of {\pimen} with the G dwarfs HD 97334 and HD 39587 
after scaling to match pi Men's distance and chromospheric activity level.
\\

Before incorporation in the analysis below, the representative 5\,\AA\--2.5\,$\mu$m spectrum 
is binned to a constant 1\,\AA. We accounted for the varying spectral 
resolutions of our data sources by convolving higher resolution data with a Gaussian 
kernel until overlapping spectral features matched.
The X-ray+EUV (5--912 {\AA}) stellar spectrum is particularly important to drive the atmospheric
escape, as both hydrogen and oxygen atoms absorb at these wavelengths. 
Our reconstruction leads
to a stellar irradiation of {\pimenc} at these wavelengths of 1350 erg cm$^{-2}$ s$^{-1}$.
\\

To test the impact of some of the above choices on the atmospheric modeling
of {\pimenc}, 
we additionally produced two alternative reconstructed spectra. 
We refer to them as the \textit{high} and \textit{low} stellar spectra, 
because they bracket the integrated EUV flux of our reference implementation. 
The X-ray+EUV flux at the planet's orbital distance for the \textit{high} spectrum is 
1810 erg cm$^{-2}$ s$^{-1}$, 
as in \citet{kingetal2019}, whereas for the \textit{low} spectrum the 
flux is 1060 erg cm$^{-2}$ s$^{-1}$, based on the estimates by \citet{franceetal2018}.
The discrepancy between these two works' EUV estimates occur because {\pimen} 
is observed to have similar chromospheric emission to the Sun \citep{franceetal2018}, 
but 3.3$\times$ higher coronal emission than the Sun \citep{kingetal2019}. 
Coronal and chromospheric emission both contribute to the unobserved EUV spectral range.

\section{Search for Transit of $\pi$\,Men\,c} \label{sec:transits}

\begin{figure}
    \centering
    \includegraphics[width=\textwidth]{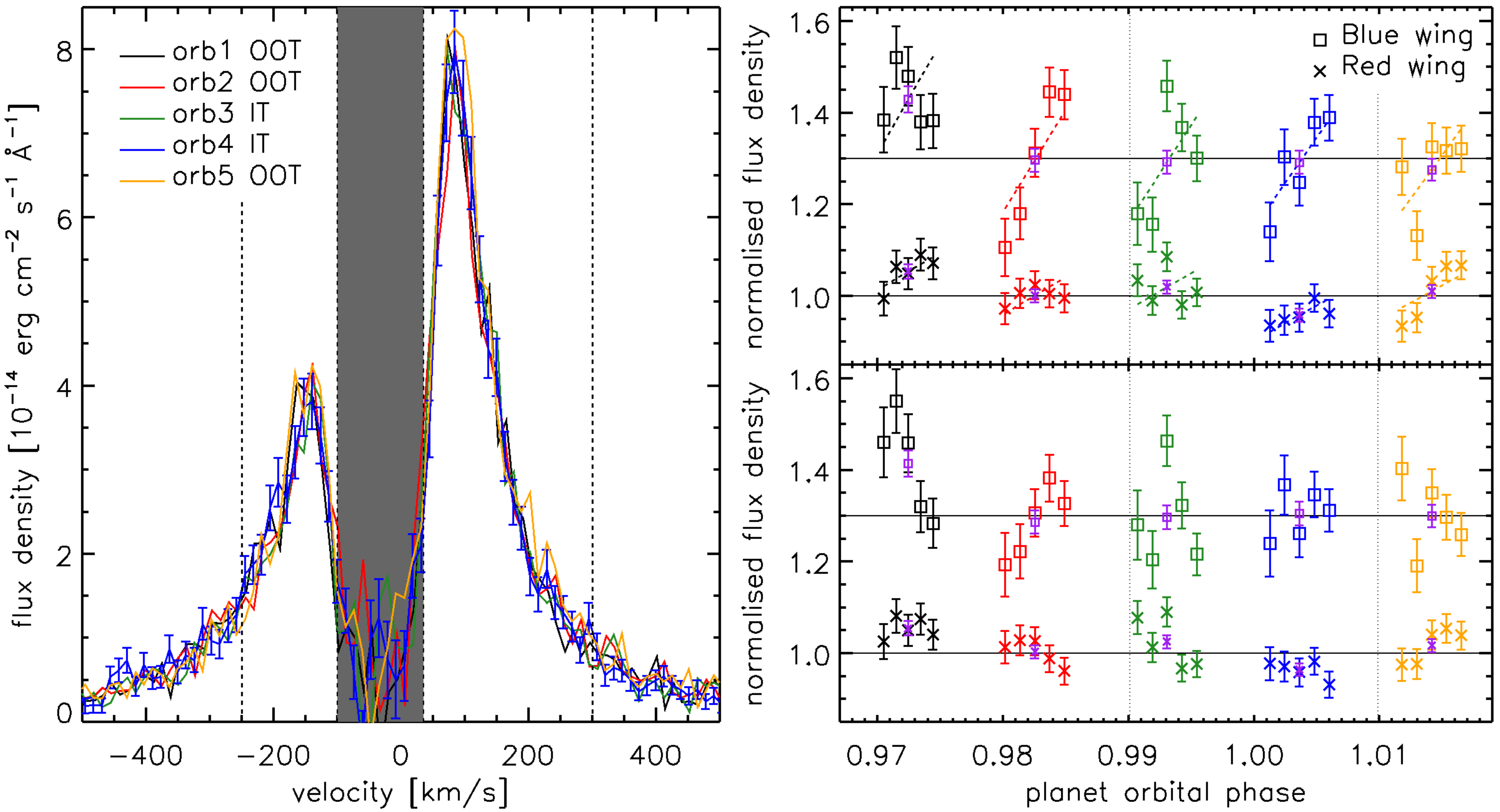}
    \caption{\textit{Left:} Comparison between the {\it calstis} extracted and 
    calibrated Ly$\alpha$ spectra of $\pi$\,Men obtained for each of the 5 HST 
    observations (orbits). The first, second, and fifth HST observations were conducted out-of-transit 
    (OOT), while the third and fourth were in-transit (IT). 
    For clarity, the uncertainties have been drawn just for the fourth observation. 
    The gray shaded area indicates the spectral region excluded from the analysis, 
    because heavily contaminated by ISM absorption and geocoronal airglow emission. 
    The dashed vertical lines mark the spectral regions in the blue and red wings 
    considered for the analysis. \textit{Top-Right:} Ly$\alpha$ light curve 
    obtained integrating across the blue (square; $-$250 to $-$100\,km/s; 
    rigidly shifted upwards by 0.3) and red (cross; +35 to +300\,km/s) wings 
    before correcting for the breathing effect. Colors are as in the left panel 
    and the purple symbols indicate the average for each HST observation. The 
    polynomials (straight lines) used to correct for the breathing effect are 
    shown by dashed lines. 
    The black horizontal lines show the modelled TESS transit light curve; its
    shape (transit depth $\sim$3$\times$10$^{-4}$) is not discernible at the 
    scale of the graph.
    \textit{Bottom-Right:} Same 
    as top-right, but after correcting for the breathing effect. No planetary 
    absorption signal is detected.}
    \label{fig:transit}
\end{figure}

A visual inspection of the extracted and calibrated STIS Ly$\alpha$ fluxes indicates that there is no clear planetary absorption signature, both in the red and blue line wings 
(Fig.~\ref{fig:transit}, left panel); fluxes obtained between $-$90 and +35\,km/s from the 
line center are strongly affected by ISM absorption and geocoronal airglow emission. 
Before performing a deeper analysis and eventually assigning upper limits on the non-detection, 
it is necessary to look for and correct for the breathing effect, which is known to affect 
most STIS observations obtained with a narrow slit 
\citep[e.g.,][]{ehrenreichetal2015,bourrieretal2018}. 
We assume that the amplitude of the breathing effect is a function of the HST orbital phase 
and that it is repeatable across the 5 HST observations. We downloaded from MAST the 
calibrated 2-dimensional spectra providing information on the photon arrival time (``\_tag.fits'' files) 
and split each HST observation in 5 sub-exposures of equal exposure time. 
From each sub-exposure image, we extracted the stellar spectra using a slanted 
extraction box with an aperture of 20 pixels and the background employing an 
identical extraction box, but shifted upwards by 100 pixels 
(see Fig.~\ref{fig:imageextraction}, Appendix \ref{ref:specextraction}). 
We then removed the relative background from the stellar spectra, 
phased each sub-exposure with HST's orbit, and finally looked for repeatable 
trends in the fluxes integrated between +35 and +300 km/s, which is the region of 
the observed {\lalpha} line with the highest flux, thus highest signal-to-noise. 
For the analysis of the breathing effect, 
we considered just the observations obtained in the second, fourth, and fifth HST observation 
because the first HST orbit is notoriously affected by additional systematics and the third 
observation has been partially obtained during the planetary ingress. We modelled the breathing 
effect as a polynomial of varying order, selecting the one that minimises the Bayesian Information 
Criterion: BIC\,=\,$\chi^2$+$k$\,$\log{N}$, where $k$ is the number of free parameters and 
$N$ is the number of data points. We finally obtained that the breathing effect is best 
described by a first order polynomial (Fig.~\ref{fig:transit}, top-right panel).
\\

We applied the same correction for the breathing effect to the whole {\lalpha}
fluxes and looked for the planetary transit signature in the light curves obtained from 
integrating across the blue ($-$250 to $-$100\,km/s) and red (+35 to +300\,km/s) line wings, 
without finding any (Fig.~\ref{fig:transit}, bottom-right panel). 
We place an upper limit on the size of the planet's H {\sc i} atmosphere by 
fitting a transit model of an opaque sphere to the {\lalpha} light curve using an MCMC technique. 
We use the \texttt{batman} package \citep{kreidberg2015} with transit parameters 
from \citet{gandolfietal2018}, and uniform limb darkening parameters. 
We find 1$\sigma$, 2$\sigma$ and 3$\sigma$ upper limits to the size of the 
planet at {\lalpha}
relative to the star $R_{Ly\alpha}$/$R_{\star}$=0.13, 0.19 and 0.24
in the [$-$215,$-$91] km/s velocity range. 
Similarly, we find
1$\sigma$, 2$\sigma$ and 3$\sigma$ upper limits 
in the [$+$57,$+$180] km/s range of 0.12, 0.16 and 0.20.

\section{Model-based intepretation} \label{sec:interpretation}

We built a model of {\pimenc}'s upper atmosphere to aid in the interpretation 
of the H {\sc i} {\lalpha} observations. 
It adds to the existing literature on hydrodynamic-photochemical models for
solar system terrestrial planets \citep[e.g.][]{kastingpollack1983,zahnlekasting1986,
chassefiere1996,tianetal2008}
and to recent investigations of CO$_2$- and H$_2$O-rich 
exoplanets \citep[e.g.][]{tian2009,johnstoneetal2018,guo2019}. 
The model solves the hydrodynamics equations for the escaping atmosphere 
considering photochemistry at pressures
$p$$\lesssim$1 dyn cm$^{-2}$ (=1 $\mu$bar, which defines the model's lower boundary) and radial distances from the planet center
$r$/$R_{\rm{p}}$=1--10. 
Our methods build upon published work for hot Jupiters and ultra-hot Jupiters 
\citep{garciamunoz2007b,garciamunozschneider2019}. 
The baseline scenario is an atmosphere whose bulk composition is dominated by 
H$_2$/H$_2$O. 
Although other compositions that include C-bearing gases
such as CO, CO$_2$ and CH$_4$ are certainly possible, we refer to the literature
\citep[e.g.][]{tian2009,johnstoneetal2018} and leave for follow-up work the investigation
of more complex hydrogen-oxygen-carbon compositions.
Our hydrogen-oxygen chemical network includes 
the neutrals H$_2$, H, H$_2$O, OH, O, O$_2$; 
the ions H$_2^+$, H$^+$, H$_3^+$, H$_2$O$^+$, H$_3$O$^+$, 
OH$^+$, O$^+$, O$_2^+$; and electrons.
They participate in 79 processes, including 14 for photodissociation/-ionization
\citep{garciamunozetal2005,garciamunoz2007a,garciamunoz2007b}. 
The energy equation includes the usual terms for
advection of total enthalpy (including chemical internal energy), thermal conduction, 
diffusive transport of enthalpy and gravitational work.
It also includes 
 radiative contributions from deposition of stellar X-ray-EUV-FUV energy 
by absorption of neutrals, 
H$_3^+$ cooling in the IR \citep{milleretal2013}, 
{\lalpha} emission excited by 
electron collisions with hydrogen \citep{black1981}, 
O($^3P$) emission at 63 and 147 $\mu$m \citep{bates1951,bankskockarts1973}, 
and H$_2$O and OH cooling through IR ro-vibrational bands \citep{hollenbachmckee1979}. 
We do not solve the problem of diffuse radiation and, 
as an upper limit to the energy radiated away by the gas, 
it is assumed that all photons from the recombinations H$^+$$+$e$\rightarrow$H+$h\nu$ and
O$^+$$+$e$\rightarrow$O+$h\nu$ are lost. 
We avoid prescribing efficiencies in the conversion from
deposited stellar energy to actual atmospheric heating. 
The model aims to explore how the mass loss rate varies with the bulk atmospheric
composition, predict the gas velocities in the upper atmosphere and the prevalent
form for each atom.
\\

{\pimenc}'s bulk composition is very uncertain. 
Thus we consider a variety of compositions for the planet's lower atmosphere that
enter into our model as boundary conditions
at the $p$=1 dyn cm$^{-2}$ level.
Effectively, we explore bulk compositions that range from 100{\%} water to 100{\%} hydrogen. 
It is assumed that there
are no bottlenecks preventing water or its dissociation products from reaching 
the upper atmosphere. 
Indeed, water condensation is unlikely for the high temperatures expected in the
lower atmosphere, which must be consistent with an  
equilibrium temperature $T_{\rm{eq}}$$\sim$1150 K.
Also, separation by mass plays a minor role for the eddy mixing considered here
(parameterized through the coefficient $K_{\rm{zz}}$=10$^8$ cm$^2$s$^{-1}$).
If a homopause exists, its location can be estimated by equating the eddy and molecular
diffusion coefficients of the relevant gases  \citep{garciamunoz2007a}. 
Assuming these are H and O   
and a local temperature $\sim$$T_{\rm{eq}}$, we estimate that the homopause occurs at 
$p$$\sim$0.2 dyn cm$^{-2}$. 
At this level, the bulk velocity of the gas is on the order of m/s, which is larger than the 
corresponding eddy or molecular diffusion velocities by  1--2 orders of magnitude. 
In summary, 
eddy diffusion ensures that gravitational settling of the heavier gases is inefficient 
below the $p$$\sim$1 dyn cm$^{-2}$ level, 
and the bulk gas velocity has a similar effect at higher altitudes. 
\\

We ran a few preliminary models with various amounts of water prescribed at the lower 
boundary that revealed that water dissociates very rapidly at the 
$p$=1 dyn cm$^{-2}$ level. 
This is consistent with the findings reported by 
\citet{guo2019} for H$_2$O-rich atmospheres. 
The leading mechanisms for H$_2$O reformation in our chemical network are 
OH+H$_2$$\rightarrow$H$_2$O+H and OH+H+H$\rightarrow$H$_2$O+H, that can barely compete with 
H$_2$O photolysis (rate coefficient 
$J_{\rm{H_2O}}$=2.7$\times$10$^{-3}$ s$^{-1}$). 
Motivated by this, we opted to prescribe the chemical composition 
at the lower boundary through the volume mixing ratios (vmrs)
of hydrogen ($x^{\rm{LB}}_{\rm{H}}$) and oxygen ($x^{\rm{LB}}_{\rm{O}}$) atoms, 
such that $x^{\rm{LB}}_{\rm{H}}$+$x^{\rm{LB}}_{\rm{O}}$$\approx$1. 
For all the other gases, we assumed photochemical equilibrium and
extrapolated their vmrs from the model cell immediately above the lower boundary.
The simulations confirmed that the overall chemistry is driven by 
the H and O atoms, and that the other gases are much less abundant and 
rapidly reach equilibrium with them. 
We explored values of $x^{\rm{LB}}_{\rm{O}}$ from 1/3 (i.e. atmospheric composition
consistent with 100{\%} water) to 0 (i.e. no water in the atmosphere).
Or equivalently, $x^{\rm{LB}}_{\rm{H}}$ from 2/3 to 1.
For consistency, we normalized all the vmrs at the lower boundary 
so that their summation is exactly one.
\\

\begin{figure*}[h]
   \centering
   \includegraphics[width=9.cm]{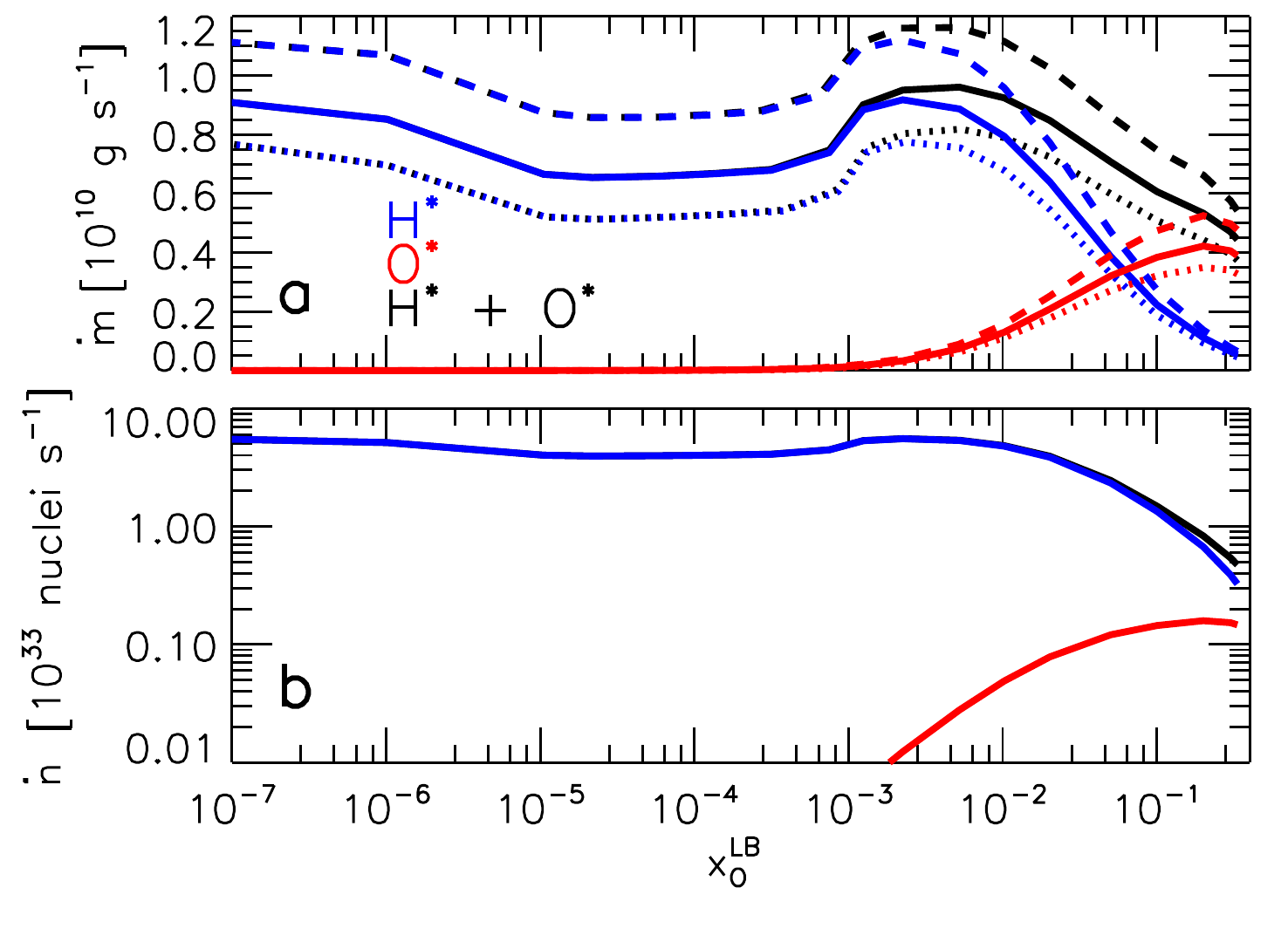}
    \label{massloss_fig}
      \caption{ 
      Loss rates for a range of bulk atmospheric compositions as 
      dictated by $x^{\rm{LB}}_{\rm{O}}$ (see text). a) Mass; b) Nuclei.
      In a) and b),
      solid lines correspond to model solutions based on our reference
      reconstruction of the stellar spectrum.
      In a), 
      dashed and dotted lines correspond to solutions for the 
      \textit{high} and \textit{low} stellar spectra, respectively.
      The loss rates differ by about $\pm$30\%, which is consistent with the difference in the
      integrated X-ray+EUV stellar fluxes.
      }
   \end{figure*}

Figure \ref{massloss_fig}a shows that the predicted mass loss rates for the
whole atmospheric gas are $\dot{m}$$\sim$4$\times$10$^9$--10$^{10}$ g s$^{-1}$
(over a solid angle $\pi$) when the reference stellar spectrum is implemented.
The mass loss rates are enhanced (diminished) when the \textit{high} (\textit{low}) 
stellar spectra are implemented, as expected. 
They are consistent with the 1.2$\times$10$^{10}$ g s$^{-1}$ 
quoted by \citet{gandolfietal2018} on the basis of the hydrogen-atmosphere models developed 
by \citet{kubyshkinaetal2018}.
These rates are moderately sensitive to the prescribed $x^{\rm{LB}}_{\rm{O}}$  
(our proxy for H$_2$/H$_2$O partitioning in the bulk atmosphere)  
even though there is a difference in the gas molecular weight by a factor of up to 6 between models.
As expected, the fractionation in the atmosphere between the hydrogen and oxygen atoms
is minor. 
Indeed, the factor 16($\dot{m}_{\rm{H}^*}$/$\dot{m}_{\rm{O}^*}$)/([H$^*$]/[O$^*$])$^{\rm{LB}}$ 
for the relative loss of hydrogen and oxygen nuclei (H$^*$, O$^*$) with respect to their
abundances at the lower boundary (brackets stand for number densities) remains in the range 1--1.1 for all cases. 
In particular, for $x^{\rm{LB}}_{\rm{O}}$=1/3 we find that 
16($\dot{m}_{\rm{H}^*}$/$\dot{m}_{\rm{O}^*}$)/([H$^*$]/[O$^*$])$^{\rm{LB}}$$\approx$1.1, 
which means that the atmosphere loses 2.2 hydrogen nuclei per oxygen nucleus, thereby 
resulting in its long-term oxidization. 
A consequence of the moderate variation in $\dot{m}$ between models is that the 
loss rate of nuclei $\dot{n}$ varies significantly with $x^{\rm{LB}}_{\rm{O}}$. 
As $x^{\rm{LB}}_{\rm{O}}$ increases (and therefore the mean molecular weight of the 
atmosphere increases too), fewer nuclei are lifted off the 
planet's gravitational potential and the planet develops a less extended atmosphere.
Figure \ref{massloss_fig}b quantifies the loss rates for
H$^*$ and O$^*$ nuclei.
\\

For models with $x^{\rm{LB}}_{\rm{O}}$=0 (black curves), 
2$\times$10$^{-2}$ (red), 
10$^{-1}$ (green) and 1/3 (blue), 
Figs. \ref{panel_fig}a--d show the profiles of: (a) velocity; 
(b) temperature; 
(c) number density of the whole gas and H atoms; 
(d) number density of O atoms and ionization fractions 
$x_{\rm{H}^+}$/$x_{\rm{H}}$ and $x_{\rm{O}^+}$/$x_{\rm{O}}$.  
Velocities $\sim$10 km/s far from the planet are established in all cases. 
At any given pressure level 
the velocities are typically lower when $x^{\rm{LB}}_{\rm{O}}$ is higher (a), 
which has a direct impact on the temperatures through adiabatic cooling (b).  
The number density of the whole gas decays more rapidly for the cases with larger 
$x^{\rm{LB}}_{\rm{O}}$ (c)  
because in the nearly-hydrostatic atmosphere the scale height is smaller for them. 
This affects how extended the upper atmosphere becomes. 
The H atom profiles (c) are directly affected by this, but also by how close to the 
planet the transition between H and H$^+$ occurs (d). 
The H$^+$/H partitioning is controlled by photoionization and the reverse process
of radiative recombination, but also by fast charge exchange 
H+O$^+$$\leftrightarrow$H$^+$+O and the fact that O atoms photoionize more readily than H atoms
because their cross sections are larger (compare $J_{\rm{O}}$=1.7$\times$10$^{-4}$ s$^{-1}$ \textit{vs.}
$J_{\rm{H}}$=5$\times$10$^{-5}$ s$^{-1}$ for unattenuated irradiation). 
Higher abundances of O atoms photoionizing more quickly (and deeper down) than H atoms   
shift the charge exchange process rightward and push the H$^+$/H transition closer to the planet. 
The result is that the H atom profiles (c) are much less extended for higher $x^{\rm{LB}}_{\rm{O}}$, 
which has direct implications for their detection. 
Through charge exchange, the H$^+$/H and O$^+$/O ratios follow nearly identical trends (d). 
Taking the number density [H]=10$^7$ cm$^{-3}$ as a reference, 
this level is reached at
$r/R_{\rm{p}}$$\sim$8, 6.9, 4.4 and 2.6, for the four cases represented in 
Figs. \ref{panel_fig}a--d in increasing order by $x^{\rm{LB}}_{\rm{O}}$. 
For comparison, 
if the two processes for charge exchange are artificially switched off in the model
for $x^{\rm{LB}}_{\rm{O}}$=1/3, the level of [H]=10$^7$ cm$^{-3}$ shifts from
$r/R_{\rm{p}}$$\sim$2.6 to 3.1. 
In summary, the assumed bulk composition at the lower boundary of the model has a major
impact on how far the H atoms extend. 
In contrast, for the three cases explored with non-zero $x^{\rm{LB}}_{\rm{O}}$, 
the reference level [O]=10$^7$ cm$^{-3}$ is reached at 
$r/R_{\rm{p}}$$\sim$2.2 (d).
\\

Figures \ref{panel_fig}e--f show 
for  $x^{\rm{LB}}_{\rm{O}}$=1/3 the number densities of various
atoms and molecules near the lower boundary (e), 
and a breakdown of the radiative terms that contribute to the energy budget (f).
It is interesting to see the spontaneous formation of moderate amounts of O$_2$ (e), 
mainly through O$+$OH$\rightarrow$O$_2$$+$H. 
Both H$_2$ (shown) and H$_2$O (not shown, but much less abundant than the other
molecules shown) are not readily formed in this oxygen-rich case, 
which supports our choice
of H and O atoms to prescribe the atmospheric composition at the lower boundary.
The energy budget over most of the upper atmosphere (f) is dominated by absorption of 
stellar X-ray+EUV radiation by H and O atoms. 
At the higher pressures investigated in the model, 
deposition of stellar FUV energy through the Schumann-Runge bands and continuum of O$_2$,  
and O($^3P$) cooling through emission at 63 $\mu$m also contribute. 

\subsection{Transit depths for H and O atoms} \label{sec:transit}

\begin{figure*}
   \centering
   \includegraphics[width=18.cm]{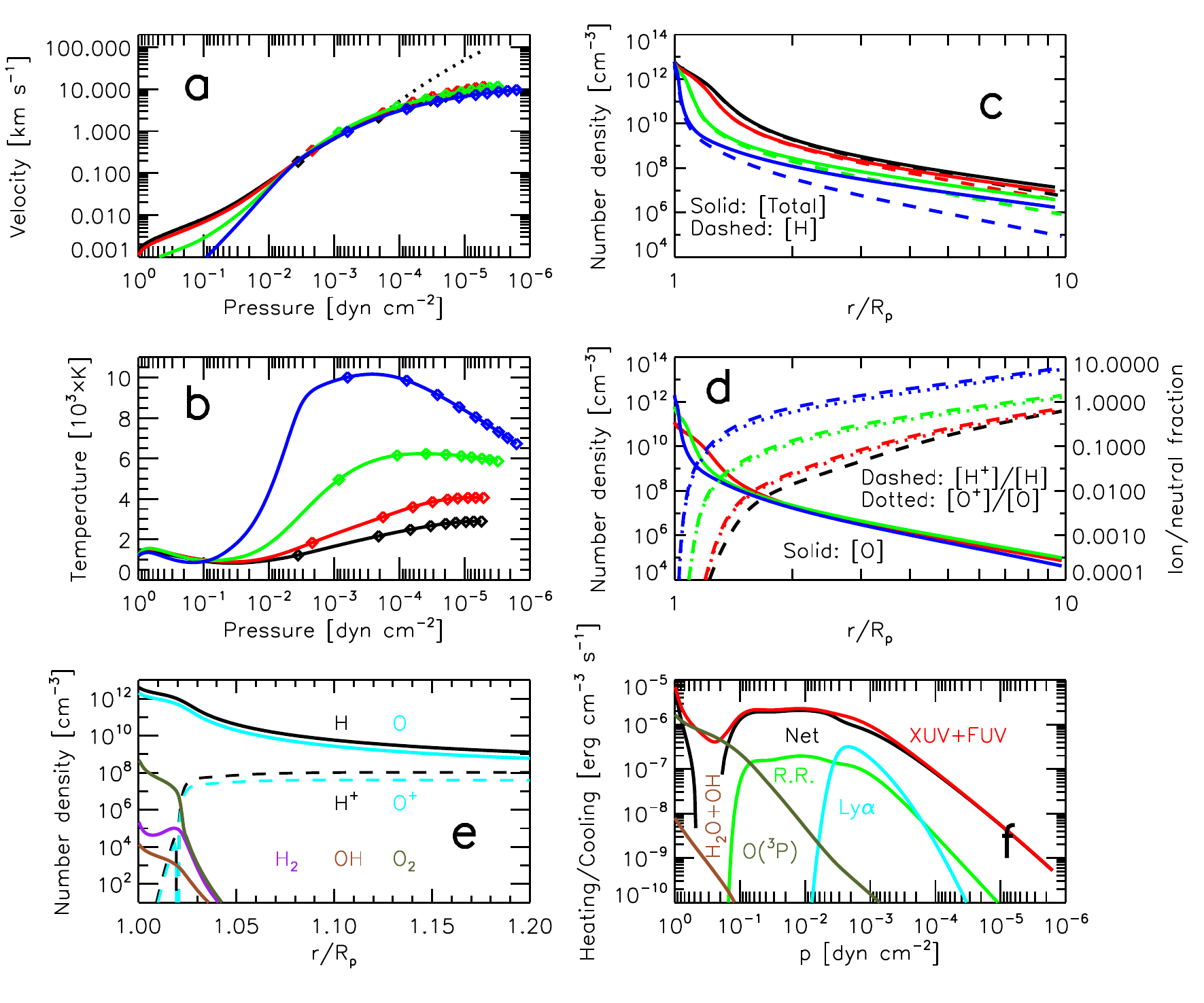}
      \label{panel_fig}
      \caption{ 
      a--d) Various atmospheric profiles for $x^{\rm{LB}}_{\rm{O}}$=0 (black), 
      2$\times$10$^{-2}$ (red), 10$^{-1}$ (green) and 1/3 (blue). 
      For $x^{\rm{LB}}_{\rm{O}}$=1/3: e) Atmospheric composition near the lower
      boundary; f) Radiative contributions to the energy budget: 
      Heating by X-ray+EUV+FUV stellar energy deposition; Cooling through various emissions (see text).
      In a), the black dotted line includes additional gas acceleration 
      (see text) on top of the $x^{\rm{LB}}_{\rm{O}}$=0 case. 
      In a--b), the symbols mark the locations for
      $r$/$R_{\rm{p}}$=1.5, 2.5, 3.5, 4.5, 5.5, 6.5, 7.5, 8.5, and 9.5.
      }
   \end{figure*}

\begin{figure*}
\centering
\includegraphics[width=9.cm]{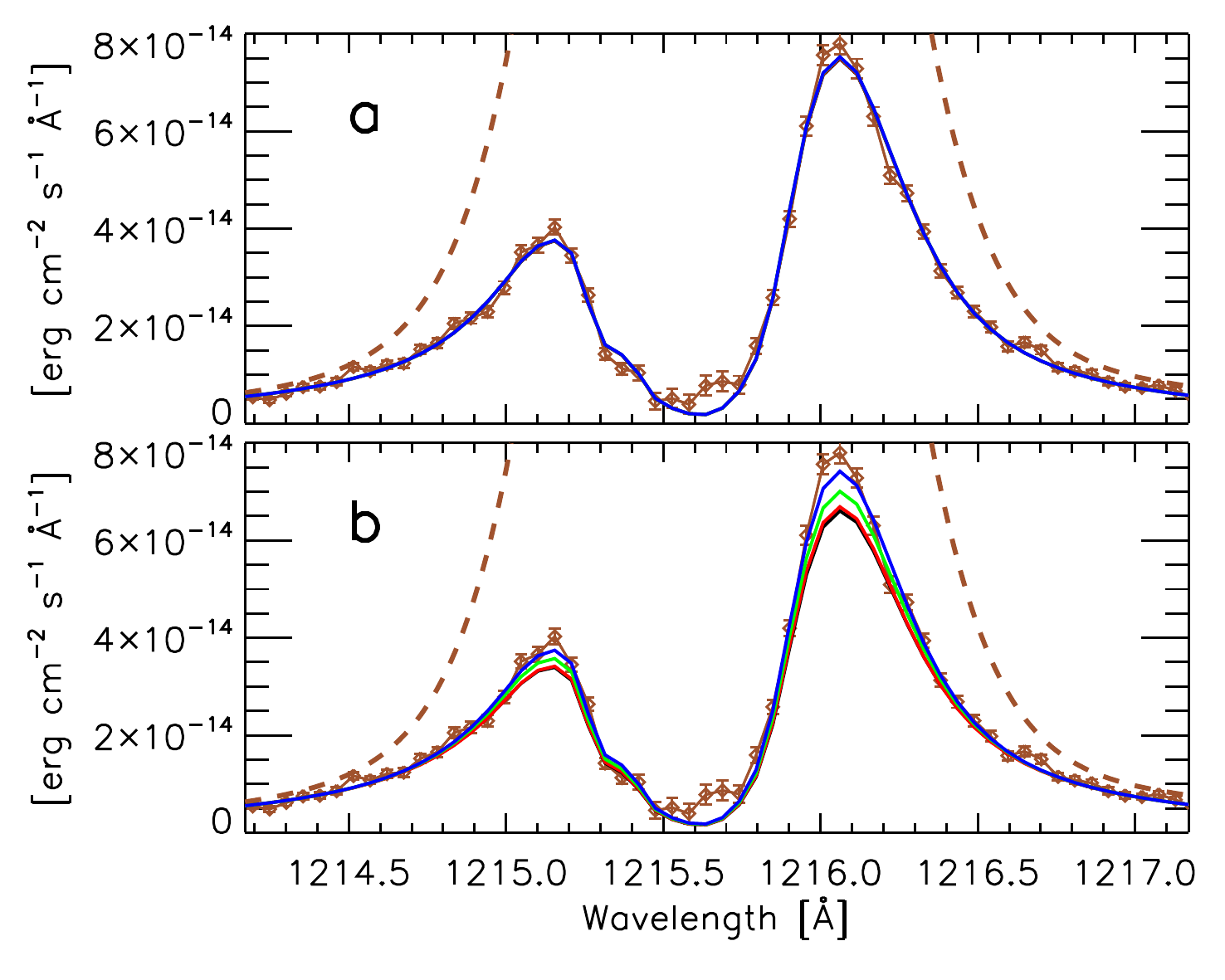}
\label{lyalpha_fig} 
\caption{
Comparison of the observed spectrum (solid brown) resulting from averaging over 
all orbits with models.
a) Based on the atmospheric profiles shown in Fig. \ref{panel_fig}.
b) Based on the atmospheric profiles with the $\psi$($r$/$R_{\rm{p}}$) transformation (see text).  
}
   \end{figure*}

We produced synthetic spectra of H {\sc i} {\lalpha} absorption at mid-transit to
compare with the observations. The absorption is described through a Voigt 
function with both thermal and natural broadening.
The transition wavelengths, probabilities and oscillator strengths 
are borrowed from the NIST Bibliographic Database \citep{kramida2010,kramidaetal2018}. 
It is assumed that the atmospheric profiles of Fig. \ref{panel_fig}  
are representative of the entire atmosphere rather than only the substellar direction. 
The gas escaping towards the star or away from it has a line-of-sight component 
that Doppler-shifts the absorption line. 
This is considered by shifting the absorption coefficient in wavelength according 
to the local line-of-sight velocity, and 
results in broader absorption spectra. 
Figure \ref{lyalpha_fig}a shows the transit spectra for  
H {\sc i} {\lalpha} absorption based on the models described above (non-brown)
 together with 
the spectrum combined from all orbits  (uncertainty bars in the measurements
have been properly reduced) (solid brown) and the reconstructed {\lalpha} line (dashed brown).  
At the scale of the plot, all four synthetic spectra are undistinguishable because most
of the absorption occurs near the {\lalpha} core which is severely affected by the ISM. 
In other words, although our model predicts a significant mass loss, 
the absorption of stellar photons by the escaping atoms 
based on the predicted velocities overlaps in wavelength with the absorption by the ISM. 
This difficulty of 1D models such as the one utilized here to 
compare with observations of H {\sc i} {\lalpha} absorption has long been known. 
\\

Indeed, 
our model does not consider the interaction of the escaping
atmosphere with radiation pressure or the stellar wind, which may accelerate 
the gas faster than the $\sim$10 km/s seen in Fig. \ref{panel_fig}a  
and enhance the absorption in the {\lalpha} wings
\citep[e.g.][]{ehrenreichetal2015,bourrieretal2018, tremblinchiang2013,trammelletal2014,
shaikhislamovetal2018,debrechtetal2019}. 
We explored this issue in an \textit{ad hoc} manner as follows. 
Assuming that the mass loss rate is determined below a few planetary radii, 
and is well predicted by our 1D model, we transformed
the velocity and density above that altitude in the way:
$u$$\rightarrow$$u$$\psi$, and [H]$\rightarrow$[H]/$\psi$, 
where $\psi$($r$/$R_{\rm{p}}$) is a function of the radial distance.
This transformation lacks a genuine physical basis but at the very least ensures mass conservation
if applied to all gases, and enables us to estimate by how much the gas should be
accelerated to produce detectable absorption in the {\lalpha} wings. 
We adopted $\psi$($r$/$R_{\rm{p}}$$<$2.5)$\equiv$1 (no additional acceleration near the 
planet) and $\psi$($r$/$R_{\rm{p}}$$\ge$2.5)=1+($r$/$R_{\rm{p}}$$-$2.5) (a linear
increase in velocity over our hydrodynamic-photochemical model predictions beyond $r$/$R_{\rm{p}}$=2.5). 
The resulting spectra, Fig. \ref{lyalpha_fig}b, 
show clear differences between model predictions, 
with the models having smaller $x^{\rm{LB}}_{\rm{O}}$ showing stronger absorption
that can be ruled out by our measurements.  
Unfortunately, our non-detection of H {\sc i} {\lalpha} absorption cannot distinguish
between the scenario of no extra acceleration represented by Fig. \ref{lyalpha_fig}a
and the scenario of possibly extra acceleration for an atmosphere with significant 
amounts of oxygen represented by $x^{\rm{LB}}_{\rm{O}}$$\ge$10$^{-1}$ in 
Fig. \ref{lyalpha_fig}b. 
Inversely, the exercise confirms that the loss of hydrogen
atoms is massive enough in the models with lower $x^{\rm{LB}}_{\rm{O}}$ 
to produce detectable H {\sc i} {\lalpha} absorption provided that 
the atoms are further accelerated at radial distances $r$/$R_{\rm{p}}$$>$2.5
to velocities a few times higher than predicted by our 1D model. 
Such a possibility motivates the current observations of {\pimenc} and 
future attempts for other small exoplanets. 
\\

Absorption by the O {\sc i} triplet at 1302-1306 {\AA} has been reported for 
the hot Jupiters HD 209458 b \citep{vidalmadjaretal2004} and HD 189733 b
\citep{benjaffelballester2013}. 
We produced O {\sc i} absorption spectra for {\pimenc}, 
and found that the transit depths are comparable 
in our cases with non-zero $x^{\rm{LB}}_{\rm{O}}$, a finding that simply reflects
that the atom number density profiles are similar (Fig. \ref{panel_fig}d). 
The transit depths at the core of the strongest of the triplet components
(at 1302 {\AA}),
and where the strongest absorption will occur, 
are on the order of 10\% for the standard atmospheric profiles, 
and 1.5--2\% for the transformed profiles.  
The drop in the transit depth for the transformed profiles is due to 
the fact that in this case the line wings absorb comparably to the line core. 
In practice, an observation that integrated over wavelengths bracketing the line core will 
result in smaller transit depths than the 10\% and 1.5--2\% estimated here. 

\section{Discussion and perspective}

{\pimenc} is probably one of the best targets to investigate  
what small exoplanets are made of. 
Taking the bulk density as a proxy for bulk composition, 
GJ 436 b ($\rho_{\rm{p}}$=1.8 g cm$^{-3}$)
is the planet most similar to {\pimenc} ($\rho_{\rm{p}}$=2.82 g cm$^{-3}$) for which 
H {\sc i} {\lalpha} absorption has been detected. 
\citet{loydetal2017} did not find evidence for C {\sc ii} or Si {\sc iii} at GJ 436 b, 
which suggests that these atoms occur in trace amounts in the atmosphere of this 
warm Neptune-size planet.
Using a hydrodynamic-photochemical model similar to ours, they also 
estimate that GJ 436 b loses mass at a rate of 3.1$\times$10$^9$ g s$^{-1}$. 
This is smaller but comparable to our estimate for {\pimenc}
(4$\times$10$^9$--10$^{10}$ g s$^{-1}$). 
Even accounting for the larger stellar size of {\pimen} (1.1$R_{\rm{\sun}}$ \textit{vs.}
0.46$R_{\rm{\sun}}$), it was reasonable to expect that if {\pimenc}'s atmosphere is
hydrogen-dominated, there would be evidence for an extended atmosphere in our HST/STIS
measurements. 
Alternatively, and assuming that the physics implemented in these models is 
relatively complete, it is fair to argue that our non-detection suggests that 
{\pimenc}'s atmosphere is not hydrogen-dominated. 
Indeed, our models show that compositions consistent with H$_2$O, and possibly other heavy 
molecules such as CO$_2$ \citep{tian2009}, will result in 
reduced number densities of H {\sc i} in {\pimenc}'s upper atmosphere even if 
the mass loss rate remains high. 
This occurs also if hydrogen is relatively abundant at the base of the upper atmosphere. 
\\

\begin{figure*}[h]
   \centering
   \includegraphics[width=11cm]{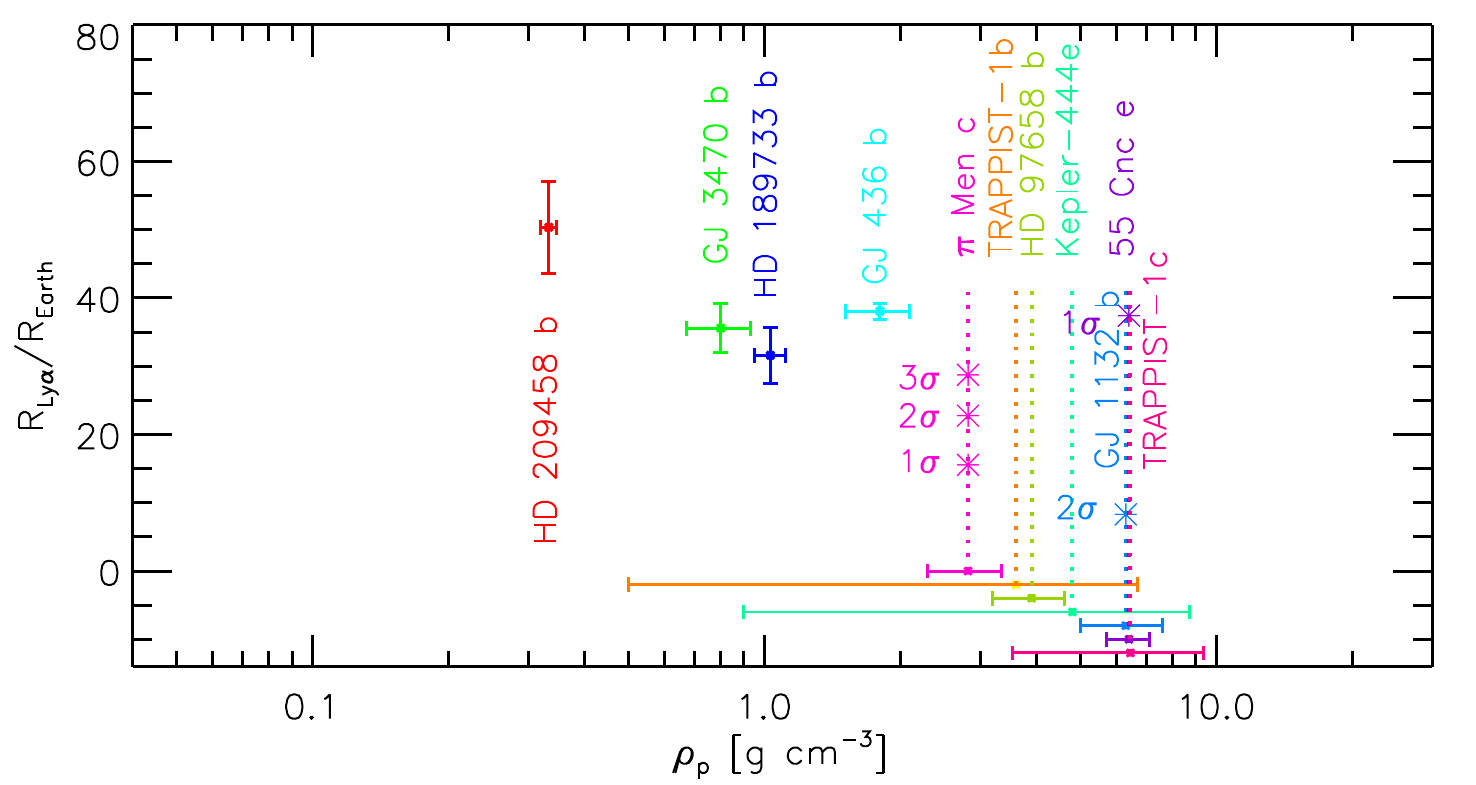} \\
   \includegraphics[width=11cm]{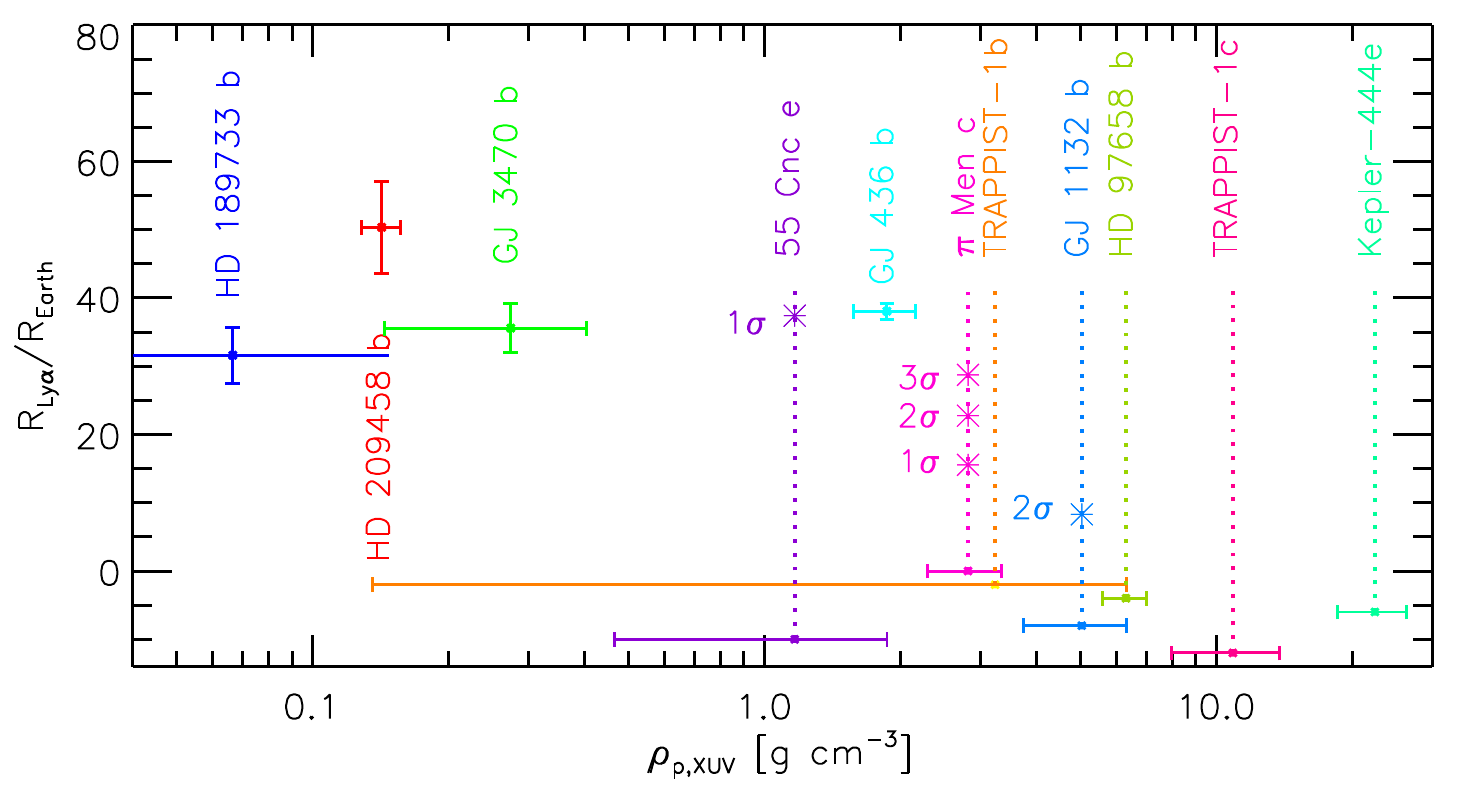}   
     \label{compo_fig}
      \caption{ \textit{Top}. For planets for which
      H {\sc i} {\lalpha} absorption measurements have been attempted, effective 
      radius at {\lalpha} normalized to Earth's radius \textit{vs.} planet density.
      For the four planets with clear detections, 
      $R_{Ly\alpha}$/$R_{\earth}$=$\sqrt{TD}$($R_{\star}$/$R_{\earth}$), 
      where $TD$ is the velocity-dependent transit depth quoted in the original references.
      For the uncertainties in $R_{Ly\alpha}$/$R_{\earth}$, we consider only the
      1$\sigma$ uncertainties in the transit depths.  
      For {\pimenc} and the other more dense planets, the measurements are
      consistent with no absorption, and we assign an arbitrary 
      $R_{Ly\alpha}$/$R_{\earth}$$<$0. 
      We omit error bars for $R_{Ly\alpha}$/$R_{\earth}$ in such cases because 
      they do not 
      necessarily reflect the planet size but other effects such as stellar variability. 
      As exceptions, we show the 1, 2 and 3$\sigma$ upper limits to 
      $R_{Ly\alpha}$/$R_{\earth}$ for {\pimenc}, 
      the 2 $\sigma$ upper limit for GJ 1132 b, and the 1 $\sigma$ upper limit for 
      55 Cnc e.
      Table \ref{planetsize_table} in Appendix \ref{ref:planetsizelyalpha} summarizes the reference sources to prepare 
      this plot.
      The diagram suggests a transition at 2--3 g cm$^{-3}$ tentatively connected to
      atmospheric composition. Hydrogen-dominated atmospheres become more extended and
      easier to detect than non-hydrogen-dominated atmospheres. 
      Amongst the planets represented here, the transition 
      is bracketed by GJ 436 b and {\pimenc}. 
      \textit{Bottom}. Similar to the above, but using the irradiation-corrected 
      density $\rho_{\rm{p,XUV}}$. 
      In the uncertainties of $\rho_{\rm{p,XUV}}$, 
      we omit the uncertainties associated with the reconstructed X-ray+EUV stellar flux      
      $F_{\rm{XUV}}$.
      Representing $R_{Ly\alpha}$/$R_{\earth}$ \textit{vs.} $\rho_{\rm{p,XUV}}$
      generally 
      confirms that less dense planets seem more prone to developing extended 
      atmospheres. 
      }
   \end{figure*}

The physics of atmospheric escape is complex, and the planet bulk density is just one of
a number of factors that play a role in it.
We summarize in Fig. \ref{compo_fig} 
previous detections and non-detections of H {\sc i} {\lalpha} absorption, 
while looking for a dependence of the measured sizes with the planets' density.  
In the Top panel 
(effective planet size at {\lalpha} $R_{Ly\alpha}$/$R_{\earth}$ \textit{vs.} planet density
$\rho_{\rm{p}}$) 
{\pimenc} sits between a group of planets for which the atmospheric composition is 
likely dominated by H$_2$/He ($\rho_{\rm{p}}$$\leq$1.8 g cm$^{-3}$)
and H {\sc i} {\lalpha} absorption has been detected, 
and another group for which other atmospheric 
compositions with large amounts of heavy molecules are possible
($\rho_{\rm{p}}$$\geq$2.8 g cm$^{-3}$) and no firm detection
of H {\sc i} {\lalpha} absorption has been reported.  
The Bottom panel is constructed in a similar way, but 
utilizes instead the irradiation-corrected density
$\rho_{\rm{p,XUV}}=\rho_{\rm{p}}({F^{\pi\;\rm{Men\;c}}_{\rm{XUV}}}/{F_{\rm{XUV}}})$
(Appendix \ref{ref:planetsizelyalpha}) to account for the theoretical
expectation that 
stronger irradiation potentially compensates for higher planet densities.
Both panels are qualitatively consistent and suggest two different behaviors 
in the planet sizes at {\lalpha} with a transition at $\rho_{\rm{p}}$=2--3
g cm$^{-3}$. 
We propose that Fig. \ref{compo_fig} may actually reflect a transition in the 
bulk composition of the planets. 
Within the transition region, GJ 436 b and {\pimenc} would 
represent hydrogen-dominated and non-hydrogen-dominated planets, respectively. 
The case of 55 Cnc e is special because its density
($\rho_{\rm{p}}$=6.4 g cm$^{-3}$) is comparatively high, but 
its irradiation-corrected density is moderately low. 
Taken at face value from Fig. \ref{compo_fig}, 
the non-detection of H {\sc i} {\lalpha} absorption at 55 Cnc e
\citep{ehrenreichetal2012} is consistent with the planet lacking an atmosphere
\citep{demoryetal2016} or having an atmosphere made of heavy molecules 
\citep{angelohu2017}.
\\

The idea that Fig. \ref{compo_fig} shows two separate behaviors
can be tested by the search for 
H {\sc i} {\lalpha} absorption, but also for O {\sc i} and C {\sc ii}, in the 
atmosphere of {\pimenc} and other small exoplanets.
In addition, further modeling will help elucidate the specifics of escape for 
non-hydrogen-dominated atmospheres and
the connection between the lower and upper atmospheres, for which little work 
has been done.

\newpage

\begin{table}[h]
\caption{Data sources for the constructed 5 \AA--2.5 {$\mu$}m spectrum of {\pimen}.  
}             
\label{table:data_sources}      
\begin{centering}                          
\begin{tabular}{c c c}        
\hline                 
\hline
Wavelength & Data & Scale \\
Range (\AA) & Source & Factor \\
\hline
5-124 & \textit{a} & 3.3$\times$ \\
124-912 & \textit{a} & 2$\times$ \\
912-1143 & \textit{a} & -- \\
$^{\dagger}$1143-1438 & \textit{b} & -- \\
1212-1220 & \textit{c} & -- \\
1280-1320 & \textit{a} & -- \\
1438-3000 & \textit{d} & 0.47$\times$ \\
3000-25,000 & \textit{e} & 2$\times$10$^{-11}$ \\
\hline
\end{tabular}
\end{centering}
\tablenotetext{\dagger}{except for Ly$\alpha$ (1212-1220 \AA) and the COS detector gap (1280-1320 \AA).}
\tablenotetext{a}{the solar minimum spectrum from \citealt{woodsetal2009}, scaled in the X-ray to match predictions from \citet{kingetal2019}, 
scaled in the EUV (124--912 {\AA}) to match an average of the predictions from \citet{kingetal2019} and \citet{franceetal2018}, 
and not scaled in the FUV based on excellent agreement with the blue end of the COS spectrum (\textit{b}).}
\tablenotetext{b}{the archival COS spectra from \citet{franceetal2018}, no scaling necessary.}
\tablenotetext{c}{the Ly$\alpha$~reconstruction from this work, shifted slightly to match the flux of the surrounding COS spectrum.}
\tablenotetext{d}{the archival STIS spectrum of the G0V star HD 39587 from \citet{ayres2010}, scaled down to match the red edge of the COS spectrum.}
\tablenotetext{e}{a G0V radiative equilibrium model from \citet{pickles1998}, scaled to match {\pimen}'s optical photometry.}
\end{table}

\newpage

\newpage

\clearpage

\appendix

\setcounter{figure}{0}
\setcounter{table}{0}

\section{The {\lalpha} flux of {\pimen} and other G dwarfs \label{ref:Gdwarfs}}

\begin{figure}[ht]
    \centering
    \includegraphics[width=\textwidth]{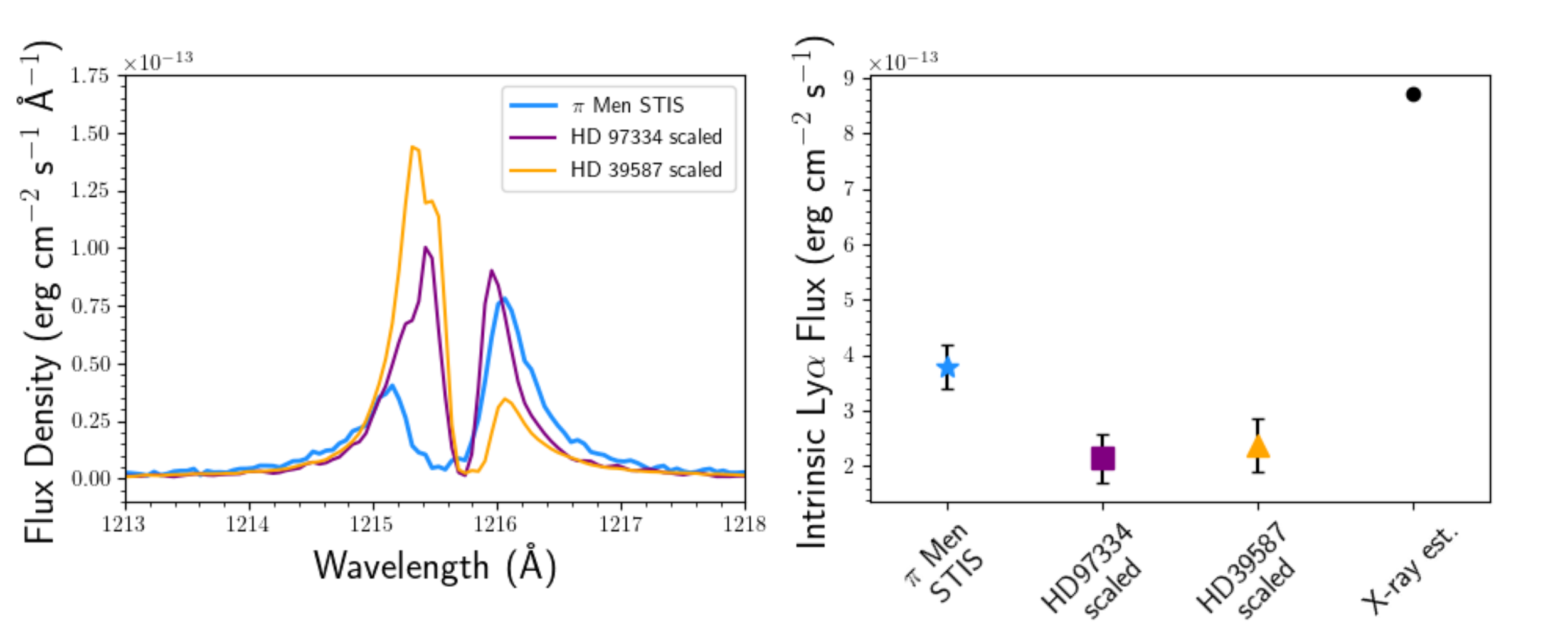}    
    \caption{\textit{Left:} 
    The co-added {\pimen} STIS spectrum (blue) is compared to the observed {\lalpha} spectra 
    of two early G dwarfs, HD 97334 (purple) and HD 39587 (orange) \citep{ayres2010}. 
    The two other stars' profiles have been convolved to match $\pi$~Men's spectral 
    resolution, scaled to match {\pimen}'s distance and further scaled 
    down to match {\pimen}'s Si {\sc iii} (1206 \AA) flux observed with COS \citep{franceetal2018}. 
    \textit{Right:} 
    The reconstructed, intrinsic {\lalpha} flux of {\pimen} from this work is compared to the 
    intrinsic {\lalpha} fluxes of HD 97334 (purple square) and HD 39587 (orange triangle)
    both from 
    \citet{woodetal2005} after applying the same scalings from the left panel. 
    Also shown is the {\lalpha} flux estimate from \cite{kingetal2019} based on 
    {\pimen}'s X-ray flux.    
    }
    \label{fig:Lya_Gdwarfs}    
\end{figure}

\newpage

\section{Information on the extraction of the spectra \label{ref:specextraction}}

\begin{figure}[h]
    \centering
    \includegraphics[width=\textwidth]{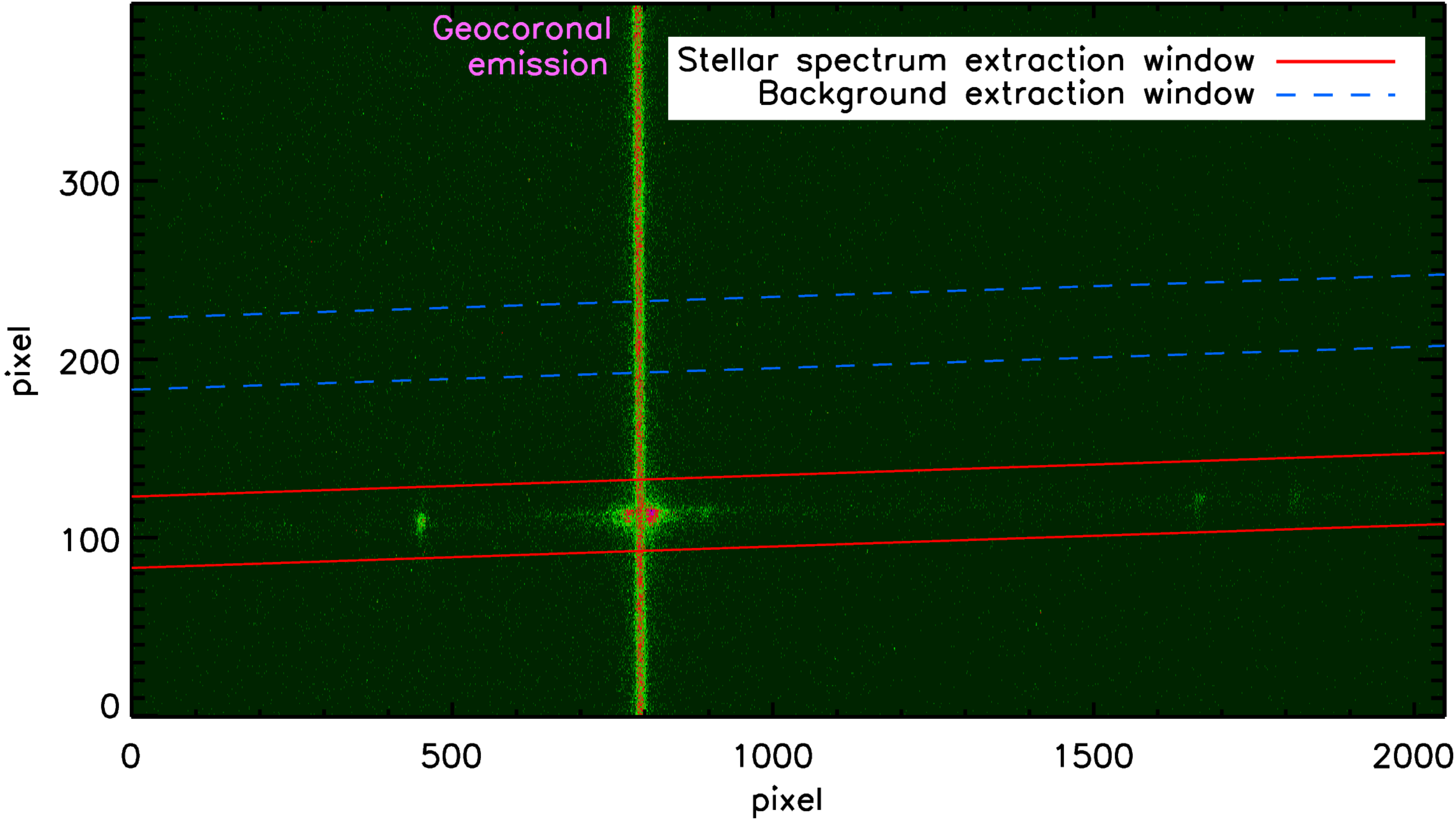}
\caption{
Image of the second HST frame in counts per second. 
The lowest counts are in dark green, while the highest counts are in purple. 
The vertical line at about 800 pixels along the $x$-axis is the geocoronal emission. 
The red solid and blue dashed lines indicate the stellar spectrum and background 
extraction boxes, respectively. The center of the stellar spectrum extraction box 
takes the function 103$+$0.012$x$, where $x$ is the pixel on the $x$-axis, 
and the extraction box has an aperture of 20 pixels, 
meaning that the total amplitude is of 40 pixels. The background extraction 
box is identical to the stellar spectrum extraction box, 
but rigidly shifted upwards by 100 pixels.    
}
    \label{fig:imageextraction}
\end{figure}

\newpage

\section{Exploring exoplanets' sizes at {\lalpha} \label{ref:planetsizelyalpha}}

For energy-limited conditions, the mass loss rate of a planet:
$$
{\dot{m}}\propto\frac{F_{\rm{XUV}}R^2_{\rm{p}}}{GM_{\rm{p}}/R_{\rm{p}}}, 
$$
where the numerator is the X-ray+EUV irradiation (wavelengths less than 912 {\AA})
received by the planet on its orbit 
over an effective area $\propto$$R^2_{\rm{p}}$, and the 
denominator is the planet's gravitational potential. 
${F_{\rm{XUV}}}/{\rho_{\rm{p}}}$ is thus a key physical parameter for
atmospheric escape.
\\

Based on the above, we define an XUV-corrected planet density: 
$$
\rho_{\rm{p,XUV}}=\rho_{\rm{p}}\frac{F^{\pi\;\rm{Men\;c}}_{\rm{XUV}}}{F_{\rm{XUV}}}, 
$$
which considers simultaneously the planet density and the effect of irradiation on
the escape. 
The choice of the X-ray+EUV irradiation for {\pimenc} as a scaling factor
ensures that $\rho_{\rm{p,XUV}}$=$\rho_{\rm{p}}$ for this planet.
Table \ref{planetsize_table} summarizes $\rho_{\rm{p}}$, $\rho_{\rm{p,XUV}}$ and
${F_{\rm{XUV}}}$ for the sample of planets considered in Fig. \ref{compo_fig}.

\begin{sidewaystable}
    \centering
\caption{Data corresponding to Fig. \ref{compo_fig}. 
TD stands for transit depth; The XUV=X-ray+EUV irradiation is at the planet's orbital
position;  
Densities and other information are taken from the NASA Exoplanet Archive 
(https://exoplanetarchive.ipac.caltech.edu/).
$\dagger$: Estimated from their Eq. (6) and quoted mass loss rate 3$\times$10$^9$ g s$^{-1}$.}             
\label{planetsize_table}      
\centering                          
\begin{tabular}{c c c c c c c c}        
\hline                 

Planet & $\rho_{\rm{p}}$ & TD & Ref. & $F_{\rm{XUV}}$ at planet & Ref. & $\rho_{\rm{p,XUV}}$\\
       &  [g cm$^{-3}$] & & &  [erg cm$^{-2}$ s$^{-1}$] & & [g cm$^{-3}$]\\
\hline                        
HD 209458 b & 0.33 & 0.15 & \citet{vidalmadjaretal2003} & 3162 & \citet{loudenetal2017} & 0.14 \\
  GJ 3470 b & 0.80 & 0.35 & \citet{bourrieretal2018} & 3938 & \citet{bourrieretal2018} & 0.27\\
HD 189733 b & 1.03 & 0.14 & \citet{lecavelierdesetangsetal2012} & 20893 & \citet{salzetal2016} & 0.07\\
GJ 436 b & 1.80 & 0.563   & \citet{ehrenreichetal2015} & 1303 & \citet{bourrieretal2018} & 1.86\\  
$\pi$ Men c & 2.82 & -- & -- & 1350 & This work & 2.82 \\  
TRAPPIST-1b & 3.60 & -- & -- & 1502 & \citet{bourrieretal2017b} & 3.23\\
HD 97658 b & 3.90 & -- & -- & 835 & \citet{bourrieretal2017a} & 6.31 \\
Kepler-444e & 4.80 & -- & -- & 289 & \citet{bourrieretal2017d} & 22.42\\
GJ 1132 b & 6.30 & -- & -- & 1689$\dagger$ & \cite{waalkesetal2019} & 5.03 \\ 
 55 Cnc e & 6.40 & -- & -- & 7413 & \citet{salzetal2016} & 1.16 \\ 
TRAPPIST-1c & 6.45 & -- & -- & 801 & \citet{bourrieretal2017b} & 10.87\\
\hline                                   
\end{tabular}
\end{sidewaystable}

\end{document}